# Topological spin transport of relativistic electron

K.Yu. Bliokh[1,2*]

[1]*Institute of Radio Astronomy, 4 Krasnoznamyonnaya st., Kharkov, 61002, Ukraine*
[2]*Bar-Ilan University, Ramat-Gan, 52900, Israel*

Semiclassical (with an accuracy of $\hbar$) motion equation for relativistic electron, which follow from the Dirac equation, are derived in this paper. We determine both the evolution equation for electron polarization, which takes the non-Abelian Berry phase into account, and Hamiltonian equations for trajectories of the particle (wave packet) center in the phase space. The equations have covariant form with respect to $U(2)$ gauge transformations and contain topological spin terms that are connected to the Berry gauge field arising during the diagonalization of the Dirac equation. The trajectory equations obtained are substantially different from the traditional ones (for instance, those that follow from the Pauli Hamiltonian) and correspond to contemporary ideas of topological spin transport of particles.

## I. INTRODUCTION

Despite its laconic form, the Dirac equation is difficult to analyze in the presence of an external electromagnetic field. The task of describing the evolution of an electron consists, in essence, of separating the negative and positive energy levels in the Dirac equation, i.e. bringing it to the block-diagonal form. This problem can be solved exactly for a free electron in the absence of an external field with the help of the Foldy-Wouthuysen transformation [1]. It can also be solved approximately in the presence of an external electromagnetic field when the momentum of the particle is small as compared with $mc$ [1]. Then the diagonalization of the Dirac Hamiltonian produces the Pauli Hamiltonian for electron with the terms of Zeeman and spin-orbital interactions.

In this paper we perform an approximate diagonalization of the Dirac equation in the presence of an electromagnetic field; however, the role of the small parameters will belong not to the electron's momentum, but to it's wavelength, or, formally, $\hbar$. Previously a semiclassical analysis of the Dirac equation was conducted in papers and books [2–9], where a number of results were obtained regarding the evolution of the electron spin, but not the trajectory equations with taking spin degree of freedom into account. Indeed, the trajectory is usually assumed to be equal to those of classical electron.[1)] In the present paper we derive the semiclassical motion equation both for its spin and translation freedom degrees with an accuracy of $\hbar$ (i.e. taking the interaction between these freedom degrees into account). We will use the contemporary theory of the motion of semiclassical particles with spin, which has revealed that it is necessary to take into account the Berry gauge potentials and fields induced during the diagonalization and transition to the adiabatic approximation (see, for instance, [10–20], and [24]). This theory in recent years lead to the revision of the motion equations of semiclassical non-relativistic particles in solids [10–16] and discovery of the *topological spin transport*, an example of which is the intrinsic spin Hall effect [11–14] experimentally confirmed recently [21]. In the case of relativistic particles, analogous equations were obtained theoretically and confirmed experimentally for photons [17–19,22], and also recently for electrons in an external electric field [20]. The present paper develops and generalizes the approach of paper [20] to the case of an arbitrary external electromagnetic field. Recent theory of semiclassical evolution of particle with degenerated levels and non-Abelian Berry gauge fields [16] has enabled us to solve the problem of the semiclassical evolution of a relativistic electron.

---

[*] E-mail: k_bliokh@mail.ru
[1)] The only exclusion is the book [9], Section 4.1.3, where the author makes an attempt to derive the translation motion equation with taking spin terms into account. However, the equation deduced therein seem to be incorrect, since the Berry curvature for the Dirac electron has been calculated as a curl of the Berry connection, which is incorrect in non-Abelian case (compare the expression for the Berry curvature in [9] with analogous expressions in [19,20] and Eqs. (24,25) below).

## II. GENERAL SEMICLASSICAL MOTION EQUATIONS

In this Section we derive semiclassical motion equations for a particle in the presence of a gauge field in a most general form. In the following Sections these equations will be applied to the Dirac equation and will help us to describe both the electromagnetic and Berry gauge field in a compact form [15]. Let the particle be described by a Schrödinger-type wave equation with a matrix Hamiltonian $\hat{H}(\mathbf{P},\mathbf{R},t)$, where $\mathbf{P}$, $\mathbf{R}$ and $t$ are the canonical momentum, coordinates and time respectively, and from here on all matrix operators are marked with hats. Let us introduce a new effective Hamiltonian

$$\hat{\mathrm{H}}(P^\alpha, R^\alpha) = \hat{H}(\mathbf{P},\mathbf{R},t) - P^0 c. \tag{1}$$

Here and further scalars added to matrices are assumed to be multiplied by the unit matrix of correspondent dimension, 4-vectors are denoted by Greek indices, the metric with signature $(-,+,+,+)$ is used, $R^0 = ct$, and the operator of total energy is $P^0 c = i\hbar\partial_t = -i\hbar c\partial_{R_0}$. As a result, the Schrödinger equation takes the form $\hat{\mathrm{H}}\psi = 0$, where $\psi = \psi(R^\alpha)$ is a multicomponent wave function, while representations of momentums and coordinates, and commutation relations take the form

$$P^\alpha = -i\hbar\partial_{R_\alpha}, \quad R^\alpha = i\hbar\partial_{P_\alpha}, \tag{2}$$

$$[P^\alpha, P^\beta] = 0, \quad [R^\alpha, R^\beta] = 0, \quad [R^\alpha, P^\beta] = i\hbar g^{\alpha\beta}. \tag{3}$$

Here $g^{\alpha\beta} = \mathrm{diag}(-1,1,1,1)$ is the metric tensor. Hamiltonian motion equations have the canonical form:

$$\dot{P}^\alpha = -\frac{i}{\hbar}[P^\alpha, \hat{\mathrm{H}}], \quad \dot{R}^\alpha = -\frac{i}{\hbar}[R^\alpha, \hat{\mathrm{H}}]. \tag{4}$$

Assume some gauge potentials on phase space, $\hat{A}_{r_\alpha}$ and $\hat{A}_{p_\alpha}$, are introduced, which shift both the momentum and coordinates in Hamiltonian (1):

$$P^\alpha \to \hat{p}^\alpha = P^\alpha - \hat{A}_{r_\alpha}, \quad R^\alpha \to \hat{r}^\alpha = R^\alpha + \hat{A}_{p_\alpha}, \tag{5}$$

with $\hat{A}_{p_0} \equiv 0$ since time is invariant. Here $\hat{p}^\alpha$ and $\hat{r}^\alpha$ are covariant momentum and coordinates, which correspond to immediately measurable momentum and coordinates of the particle (in particular they correspond to ordinary, rather than generalized, variables in classical mechanics). The gauge potentials change the commutation relations, while the Hamiltonian equations preserve their form: $\dot{\hat{p}}^\alpha = -\frac{i}{\hbar}[\hat{p}^\alpha, \hat{\mathrm{H}}], \dot{\hat{r}}^\alpha = -\frac{i}{\hbar}[\hat{r}^\alpha, \hat{\mathrm{H}}]$. In semiclassical approximation, with an accuracy of $\hbar$, these equations go to

$$\dot{\hat{p}}^\alpha = -\frac{i}{\hbar}\left[\hat{p}^\alpha,\hat{p}^\beta\right]\bullet\partial_{\hat{p}^\beta}\hat{H} - \frac{i}{\hbar}\left[\hat{p}^\alpha,\hat{r}^\beta\right]\bullet\partial_{\hat{r}^\beta}\hat{H},$$
$$\dot{\hat{r}}^\alpha = -\frac{i}{\hbar}\left[\hat{r}^\alpha,\hat{p}^\beta\right]\bullet\partial_{\hat{p}^\beta}\hat{H} - \frac{i}{\hbar}\left[\hat{r}^\alpha,\hat{r}^\beta\right]\bullet\partial_{\hat{r}^\beta}\hat{H},$$
(6)

where • stands for symmetrized product. When the commutation relations are canonical like Eqs. (3), equations (6) become classical Hamiltonian equations. Influence of the gauge field on the trajectory of motion is contained in the non-trivial commutators in Eqs. (6). For commutators of covariant variables (5) we have

$$\left[\hat{p}^\alpha,\hat{p}^\beta\right] = i\hbar\hat{F}_{r_\alpha r_\beta},\quad \left[\hat{r}^\alpha,\hat{r}^\beta\right] = i\hbar\hat{F}_{p_\alpha p_\beta},\quad \left[\hat{r}^\alpha,\hat{p}^\beta\right] = i\hbar g^{\alpha\beta} - i\hbar\hat{F}_{p_\alpha r_\beta}.$$
(7)

where $\hat{F}_{r_\alpha r_\beta} = \partial_{\hat{r}_\alpha}\wedge\hat{A}_{r_\beta} - i\hbar^{-1}\left[\hat{A}_{r_\alpha},\hat{A}_{r_\beta}\right]$, $\hat{F}_{p_\alpha r_\beta} = \partial_{\hat{p}_\alpha}\wedge\hat{A}_{r_\beta} - i\hbar^{-1}\left[\hat{A}_{p_\alpha},\hat{A}_{r_\beta}\right]$, etc. are components of the antisymmetric tensor of the gauge field (curvature) on the phase space, with $\hat{F}_{p_\alpha p_0} = \hat{F}_{r_\alpha p_0} \equiv 0$.

To describe a particle with an additional, spin freedom degree, one need to introduce the unit vector of polarization, $|\chi\rangle$. Then a classical value that corresponds to the center of semiclassical particle (wave packet) is produced from a given quantum operator by replacing the differential operators (2) with the corresponding classical values, and by convolution of its matrix with the polarization vector: $p^\alpha = \langle\chi|\hat{p}^\alpha|\chi\rangle$, $F_{r_\alpha r_\beta} = \langle\chi|\hat{F}_{r_\alpha r_\beta}|\chi\rangle$, $H = \langle\chi|\hat{H}|\chi\rangle$, etc. In what follows we denote classical parameters of the particle's center by the same characters without hats. As a result, semiclassical equations of motion of the particle's center follow from Eqs. (6):

$$\dot{p}^\alpha = -\partial_{r_\alpha}H + F_{r_\alpha r_\beta}\partial_{p_\beta}H - F_{r_\alpha p_\beta}\partial_{r_\beta}H,\quad \dot{r}^\alpha = \partial_{p_\alpha}H + F_{p_\alpha p_\beta}\partial_{r_\beta}H - F_{p_\alpha r_\beta}\partial_{p_\beta}H.$$
(8)

In spite of the fact that these are equations for classical dynamical variables, they can contain terms of the order of $\hbar$, which describe the influence of the spin degree of freedom.

## III. SEMICLASSICAL DIAGONALIZATION OF THE DIRAC EQUATION

The effective Hamiltonian (1) for the Dirac equation with an external electromagnetic field is equal

$$\hat{H}(P^\alpha, R^\alpha) = \hat{\alpha}_\beta\left(P^\beta - \frac{e}{c}\mathscr{A}^\beta(R^\alpha)\right)c + \hat{\beta}mc^2,$$
(9)

where $\hat{\beta}$ and $\hat{\alpha}^\alpha = (\hat{\alpha}^0,\hat{\boldsymbol{\alpha}})$ are Dirac matrices, with $\hat{\alpha}^0 = \hat{1}$ is the unit matrix, $\mathscr{A}^\alpha$ is the electromagnetic potential, and summation over repeated indices is understood. Formally, Hamiltonian (9) is of the form of the Hamiltonian of a free electron when one introduces momentum $\tilde{p}^\alpha = P^\alpha - \frac{e}{c}\mathscr{A}^\alpha$. However, it cannot be diagonalized by Foldy-Wouthuysen transformation for free particle [1,20], because of non-trivial commutation relations of the space components of $\tilde{p}^\alpha$: $[\tilde{p}_i,\tilde{p}_j] = i\frac{e\hbar}{c}\varepsilon_{ijk}\mathscr{H}_k$, where $\mathscr{H}$ is the external magnetic field. The matter of the fact is that relation $(\hat{\boldsymbol{\alpha}}\mathbf{P})^2 = P^2$ (here and further $P^2 \equiv \mathbf{P}^2$), crucial for Foldy-Wouthuysen transformation, fails. Instead of this, the following relation takes place for momentum $\tilde{\mathbf{p}}$: $(\hat{\boldsymbol{\alpha}}\tilde{\mathbf{p}})^2 = \tilde{p}^2 - \frac{e\hbar}{c}\hat{\Sigma}\mathscr{H}$, where $\hat{\Sigma}/2$ is the operator of spin of Dirac particle.

We reveal that if one makes a substitution $P^2 \to (\hat{\alpha}\tilde{\mathbf{p}})^2 = \tilde{p}^2 - \frac{e\hbar}{c}\hat{\Sigma}\mathcal{H}$ in the Foldy-Wouthuysen transformation for free particle, then it diagonalizes Dirac equation in semiclassical approximation with an accuracy of $\hbar$. (Semiclassical approximation is needed here to consider $\tilde{\mathbf{p}}$ and $\hbar\mathcal{H}$ as commuting values.) Thus, the modified Foldy-Wouthuysen transformation is determined by unitary operator

$$\hat{U}(\tilde{\mathbf{p}}) = \frac{\bar{E}_{\tilde{p}} + mc^2 + \hat{\beta}\hat{\alpha}\tilde{\mathbf{p}}c}{\sqrt{2\bar{E}_{\tilde{p}}(\bar{E}_{\tilde{p}} + mc^2)}}, \qquad (10)$$

where $\bar{E}_{\tilde{p}} = \sqrt{m^2c^4 + (\hat{\alpha}\tilde{\mathbf{p}})^2 c^2} \simeq E_{\tilde{p}} - \frac{e\hbar c}{2E_{\tilde{p}}}\hat{\Sigma}\mathcal{H}$, $E_{\tilde{p}} = \sqrt{m^2c^4 + \tilde{p}^2c^2}$, and all calculations are made with an accuracy of $\hbar$ in Hamiltonian. By making a substitution $\psi = \hat{U}^\dagger \psi'$ with operator (10) we have a transformation for Hamiltonian (9): $\hat{H}' = \hat{U}\hat{H}\hat{U}^\dagger$. Calculations lead to the following Hamiltonian:

$$\hat{H}'(\hat{p}^\alpha, \hat{r}^\alpha) = \hat{\beta} E_{\hat{p}} + \Delta\hat{E}(\hat{p}^\alpha, \hat{r}^\alpha) - \hat{p}^0 c. \qquad (11)$$

Here

$$\hat{p}^\alpha = P^\alpha - \frac{e}{c}\mathcal{A}^\alpha(\hat{r}^\alpha) - \hbar\hat{A}_{R_\alpha}, \quad \hat{r}^\alpha = R^\alpha + \hbar\hat{A}_{P_\alpha}, \qquad (12)$$

are covariant coordinates determined by two gauge potentials: the electromagnetic one and pure gauge potential $\hat{A}_{M_\alpha} = i\hat{U}\partial_{M_\alpha}\hat{U}^\dagger$ (with $\hat{A}_{P_\alpha} = (0, \hat{\mathbf{A}}_P)$) induced by transformation (10) on the phase space $M^\alpha = (P^\alpha, R^\alpha)$. The second term in Hamiltonian (11) equals

$$\Delta\hat{E} = -\frac{e\hbar c}{2E_{\hat{p}}}\hat{\Sigma}\mathcal{H} - \frac{ec}{E_{\hat{p}}}\hat{\mathbf{L}}\mathcal{H}, \qquad (13)$$

where, as it is shown in [10,16,23], operator $\hat{\mathbf{L}} = \hbar\hat{\mathbf{p}} \times \hat{\mathbf{A}}_P = (\mathbf{R} - \hat{\mathbf{r}}) \times \hat{\mathbf{p}}$ has the meaning of the intrinsic angular momentum of semiclassical particle. The value $\Delta\hat{E} \sim \hbar$ is the energy of the interaction of spin and intrinsic angular momentums with the magnetic field.

Hamiltonian (11) is almost block-diagonal. Cross terms, which connect the upper and lower components of the four-component wave function $\psi'$, is of the order of $\hbar$ and arise due to the non-block-diagonal structure of the gauge potential $\hat{A}_{M_\alpha}$. In adiabatic approximation (i.e. when considering the positive energy level only) the left upper $2 \times 2$ sector of Eq. (11) presents the desired Hamiltonian of the electron with an accuracy of $\hbar$. Correspondingly the wave function of the electron, $\psi^+$, is represented by two upper components of $\psi'$. By marking the left upper $2 \times 2$ sector of matrix operators by superscript "+" (with $\hat{H}'^+ \equiv \hat{H}^+$) we obtain

$$\hat{H}^+(\hat{p}^{+\alpha}, \hat{r}^{+\alpha}) = E_{\hat{p}^+} + \Delta\hat{E}^+(p^\alpha, r^\alpha) - \hat{p}^{+0}c, \qquad (14)$$

$$\hat{p}^{+\alpha} = P^\alpha - \frac{e}{c}\mathcal{A}^\alpha(\hat{r}^{+\alpha}) - \hbar\hat{A}^+_{R_\alpha}, \quad \hat{r}^{+\alpha} = R^\alpha + \hbar\hat{A}^+_{P_\alpha}, \qquad (15)$$

$$\Delta\hat{E}^+ = -\frac{e\hbar c}{2E_p}\hat{\sigma}\mathcal{H} - \frac{ec}{E_p}\hat{\mathbf{L}}^+\mathcal{H}, \qquad (16)$$

where $\hat{\sigma}$ is the vector of Pauli matrices. In Eqs. (14), (16) and further we use that in the approximation considered $\hat{p}^{+\alpha} \simeq \tilde{p}^\alpha \simeq p^\alpha$, $\hat{r}^{+\alpha} \simeq R^\alpha \simeq r^\alpha$ in terms of the order of $\hbar$ ($p^\alpha = \langle\chi|\hat{p}^{+\alpha}|\chi\rangle$ and $r^\alpha = \langle\chi|\hat{r}^{+\alpha}|\chi\rangle$ are the electron's classical momentum and coordinates). Potentials $\hat{A}^+_{P_\alpha}$, $\hat{A}^+_{R_\alpha}$ are components of the Berry gauge potential (Berry connection); it describes the non-trivial geometry and topology of the fiber bundle of Hamiltonian eigen vectors over the electron's phase space. In physical terms it is responsible for the interaction between spin and translation freedom degrees [9] or for a latent influence of the negative-energy level (see [24]). Direct calculations lead to the following components of the Berry gauge potential (compare with [7,9,20]):

$$\hat{A}^+_{P_\alpha} = \left(0, \hat{\mathbf{A}}^+_{\mathbf{P}}\right), \quad \hat{\mathbf{A}}^+_{\mathbf{P}} = \frac{(\tilde{\mathbf{p}} \times \hat{\boldsymbol{\sigma}})c^2}{2E_{\tilde{p}}\left(E_{\tilde{p}} + mc^2\right)}, \quad \hat{A}^+_{R_\alpha} = -\frac{e}{c}\left(\partial_{R_\alpha} \mathscr{A}_\beta\right)\hat{A}^+_{P_\beta}. \tag{17}$$

When one use the Hamiltonian as an operator acting on the wave function (rather than a function of the covariant position and momentum operators) it is more convenient to represent (14) as a function of canonical variables:

$$\hat{H}^+\left(P^\alpha, R^\alpha\right) = E_{\tilde{p}} + \Delta\hat{E}^+\left(\tilde{p}^\alpha, R^\alpha\right) - \tilde{p}^0 c - \hbar\hat{A}^+_{M_\beta}\dot{M}_\beta. \tag{18}$$

This form of the Hamiltonian has been obtained earlier in [7–9] and it is useful for the analysis of polarization evolution and Berry phase. The energy correction $\Delta\hat{E}^+$ is so-called 'no-name term' in [7–9,24], while the last term in Eq. (18) is accountable for $SU(2)$ holonomy described by the Berry gauge potential. By using Eqs. (17) and the motion equation of zero (classical) approximation, one can deduce that relation $\hat{A}^+_{M_\beta}\dot{M}_\beta = \hat{\mathbf{A}}\dot{\mathbf{p}}$ takes place, hence the last, topological term in Eq. (18) is determined in fact by the Berry gauge potential in $\mathbf{p}$-space only. Substitution of $\dot{M}^\alpha$ from classical motion equations with Eqs. (16), (17) leads Eq. (18) to (see [7–9])

$$\hat{H}^+ = E_{\tilde{p}} - \frac{e\hbar c}{2E_p}\hat{\boldsymbol{\sigma}}\cdot\mathscr{H} - \frac{e\hbar c^2}{2E_p\left(E_p + mc^2\right)}\hat{\boldsymbol{\sigma}}\left(\mathscr{E}\times\mathbf{p}\right) - \tilde{p}^0 c. \tag{19}$$

where $\mathscr{E}$ is the external electric field. In the non-relativistic limit Eq. (19) goes to the well-known Pauli Hamiltonian with terms of Zeeman and spin-orbit interactions. In so doing expression $-\frac{ec}{E_{\tilde{p}}}\hat{\mathbf{L}}^+\mathscr{H} - \hbar\hat{\mathbf{A}}^+_{\mathbf{P}}\dot{\tilde{\mathbf{p}}} = -e\hbar\hat{\mathbf{A}}^+_{\mathbf{P}}\mathscr{E}$ corresponds to the spin-orbit interaction term in Eq. (19) (compare with [20]). Thus, equations (18), (19) are the semiclassical generalization of Pauli Hamiltonian for the case of arbitrary energies.

## IV. ELECTRON'S MOTION EQUATIONS

The evolution of a semiclassical particle can be described by motion equations for its center and for non-Abelian (in the case of degenerated level) phase or polarization. Coordinates of the electron's center in the phase space are determined by covariant variables $\hat{p}^{+\alpha}$, $\hat{r}^{+\alpha}$, and Hamiltonian (14), whereas the evolution of its phase and polarization is determined through the action, i.e. through canonical coordinates and Hamiltonian (18) or (19). Let us consider first the evolution of the polarization.

Motion equation for the unit vector of polarization, $|\chi\rangle$, can be written out based on the Hamiltonian (18) (see [16,19]). By omitting scalar terms, which result in the appearance of usual $U(1)$ phase, for $SU(2)$ evolution of polarization we have:

$$|\dot{\chi}\rangle = i\left[-\Delta\hat{E}^+ + \hat{A}_{M_\beta}\dot{M}_\beta\right]|\chi\rangle. \tag{20}$$

One can write down the solution of this equation as

$$|\chi\rangle = \mathscr{P}\exp\left[-i\int_0^t \Delta\hat{E}^+ dt' + i\int_C \hat{A}_{M_\beta} dM_\beta\right]|\chi_0\rangle, \tag{21}$$

where $\mathscr{P}$ is the operator of chronological ordering, and $C$ is the contour, along which the electron moves in the classical phase space. In non-Abelian phase of Eq. (21) the non-Abelian Berry phase is present that corresponds to general theory [16]. In fact it is determined by the Berry gauge potential in **p**–space only: $\hat{\theta}_B = \int_C \hat{A}_{M_\beta} dM_\beta = \int_{C_\mathbf{p}} \hat{\mathbf{A}}(\mathbf{p}) d\mathbf{p}$. Here $C_\mathbf{p}$ is the contour of the evolution in **p**–space. If we introduce unit classical spin vector $\mathbf{S} = (\chi | \hat{\boldsymbol{\sigma}} | \chi)$, then Eq. (20) results in

$$\dot{\mathbf{S}} = -\left[\frac{ec}{2E_p}\mathcal{H} + \frac{ec^2}{2E_p(E_p + mc^2)}\mathcal{E} \times \mathbf{p}\right] \times \mathbf{S}. \tag{22}$$

This is known motion equation for the classical spin of the electron (so-called BMT-equation), that has been proposed for the first time in [4] and rigorously derived in [7–9].

Now let us deduce semiclassical equations of motion of the electron's center. According to Eqs. (7), (8) they are determined by commutators of covariant variables. Rather complicated but direct calculations of commutators of Eqs. (15) with Eqs. (17) give:

$$\left[\hat{r}^{+\alpha}, \hat{r}^{+\beta}\right] = i\hbar^2 \hat{F}^{\alpha\beta}, \quad \left[\hat{p}^{+\alpha}, \hat{p}^{+\beta}\right] = i\hbar \frac{e}{c} \mathcal{F}^{\alpha\beta} + i\hbar^2 \frac{e^2}{c^2} \mathcal{F}^{\alpha\gamma} \mathcal{F}^{\beta\delta} \hat{F}_{\gamma\delta},$$

$$\left[\hat{p}^{+\alpha}, \hat{r}^{+\beta}\right] = -i\hbar g^{\alpha\beta} - i\hbar^2 \frac{e}{c} \mathcal{F}^{\alpha\gamma} \hat{F}^\beta{}_\gamma. \tag{23}$$

Here $\mathcal{F}^{\alpha\beta} = \partial_{R_\alpha} \wedge \mathcal{A}^\beta$ is the tensor of the electromagnetic field with $\mathcal{F}^{\alpha\beta} = \mathcal{F}^{\alpha\beta}(\hat{r}^{+\alpha})$ in Eqs. (23), $\hat{F}^{\alpha\beta} \equiv \hat{F}^+_{P_\alpha P_\beta} = \partial_{P_\alpha} \wedge \hat{A}^+_{P_\beta} - i\left[\hat{A}^+_{P_\alpha}, \hat{A}^+_{P_\beta}\right]$ is the antisymmetric tensor of Berry gauge field (Berry curvature) in the momentum space, and in derivation of the momentums commutator in Eq. (23) Maxwell equation $\partial_{R_\alpha} \mathcal{F}^{\beta\gamma} + \partial_{R_\beta} \mathcal{F}^{\gamma\alpha} + \partial_{R_\gamma} \mathcal{F}^{\alpha\beta} = 0$ was used [25]. Similar to Eqs. (17) all topological values are expressed through Berry gauge field on **p**–space, in which only space components are non-zero, and hence it can be characterized by the dual vector $\hat{\mathbf{F}}$ ($\hat{F}^{ij} = \varepsilon^{ijk}\hat{F}_k$, $\varepsilon_{ijk}$ is the unit antisymmetric tensor):

$$\hat{\mathbf{F}} = -\frac{c^4}{2E_p^3}\left[m\hat{\boldsymbol{\sigma}} + \frac{(\hat{\boldsymbol{\sigma}}\mathbf{p})\mathbf{p}}{E_p + mc^2}\right]. \tag{24}$$

It is worth noting that commutators in Eq. (23) do not represent the sum of the electromagnetic field and Berry gauge field determined in a standard way through potentials (17) [10,14–16]. Expressions (23) essentially contain commutators of the electromagnetic potential and the Berry gauge potential (17). This is that that provides for covariant form of Eqs. (23) and points out that *Berry gauge field cannot be introduced as additive to the electromagnetic one* (only potentials are additive but non-commuting quantities). It is correct to talk only of the total gauge field in Eqs. (23), which is determined by the total gauge potential: electromagnetic plus Berry's. The total gauge field is U(2) one, whereas the proper electromagnetic potential determines U(1) field, and the Berry gauge potential determines SU(2) field.

Substituting Eqs. (14) and (23) in Eqs. (8) one can derive semiclassical motion equations of the electron. By introducing momentum, coordinates and other classical quantities for the particle's center as $\mathrm{H} = (\chi | \hat{\mathrm{H}}^+ | \chi)$, $\Delta E = (\chi | \Delta \hat{E}^+ | \chi)$, $\mathbf{F} = (\chi | \hat{\mathbf{F}} | \chi)$, etc., with

$$\mathbf{F}(\mathbf{p}, \mathbf{S}) = -\frac{c^4}{2E_p^3}\left[m\mathbf{S} + \frac{(\mathbf{S}\mathbf{p})\mathbf{p}}{E_p + mc^2}\right], \tag{25}$$

we obtain resulting motion equations:

$$\dot{p}^\alpha = -\partial_{r_\alpha} \mathrm{H} + \frac{e}{c}\mathcal{F}^{\alpha\beta}\dot{r}_\beta, \quad \dot{r}^\alpha = \partial_{p_\alpha} \mathrm{H} - \hbar F^{\alpha\beta}\dot{p}_\beta. \tag{26}$$

This is the most laconic 4-dimensional form of the main equations of the present paper. They show that the electron center trajectory is determined by two gauge fields: electromagnetic one and Berry gauge field in the momentum space. As in the case of polarization, Eq. (22), the other components of the Berry gauge field do not contribute to the motion equations. Since the Berry gauge field (25) depends on the electron's spin vector $\mathbf{S}$, the motion equations (26) form complete system together with the spin motion equation (22). In the 3-dimensional form the space components of equations (26) with substitution of Hamiltonian (14) yield

$$\dot{\mathbf{p}} = -\partial_r \Delta E + e\mathcal{E} + \frac{e}{c}\dot{\mathbf{r}} \times \mathcal{H}, \quad \dot{\mathbf{r}} = \frac{\mathbf{p}c^2}{E_p} + \partial_p \Delta E - \hbar \dot{\mathbf{p}} \times \mathbf{F}. \quad (27)$$

The topological term in the second equation has the form of the 'Lorentz force' that is caused by the 'magnetic field' $\mathbf{F}$ in the momentum space. In this representation the topological term looks exactly as in the absence of the external magnetic field [20], but in fact equations (27), being resolved with an accuracy of $\hbar$ relatively to the time derivatives, are of a rather complex form with a number of terms unknown before:

$$\dot{\mathbf{p}} = -\partial_r \Delta E + e\mathcal{E} + \frac{ec}{E_p}\mathbf{p} \times \mathcal{H} + \frac{e}{c}\partial_p \Delta E \times \mathcal{H} + \hbar \frac{e^2}{c}(\mathbf{F} \times \mathcal{E}) \times \mathcal{H} + \hbar \frac{e^2}{E_p}(\mathbf{F}\mathcal{H})\mathbf{p} \times \mathcal{H},$$

$$\dot{\mathbf{r}} = \frac{\mathbf{p}c^2}{E_p} + \partial_p \Delta E + \hbar e \mathbf{F} \times \mathcal{E} + \hbar \frac{ec}{E_p}\mathbf{F} \times (\mathbf{p} \times \mathcal{H}). \quad (28)$$

## V. SIMPLE COROLLARIES

Motion equations (28) are substantially different from the equations for canonical variables, which follow from Pauli-type Hamiltonian (18), (19). Equations (28) is of covariant form, that is, they transform covariantly under $\mathrm{U}(2)$ gauge transformations, which are connected with the freedom of the choice of a basis corresponding to double degenerated energy level. The topological terms (i.e. those that are proportional to the Berry gauge field) in the first equation (28) represent a force nonlinear with respect to the external electromagnetic field. This points at essential entanglement of the two gauge fields separated formally in Eqs. (26), (27).

Without the magnetic field in the non-relativistic limit the first equation (28) becomes evident, $\dot{\mathbf{p}} = e\mathcal{E}$, while the second one with Eq. (25) yields with an accuracy of $1/m^3c^3$

$$\dot{\mathbf{r}} = \frac{\mathbf{p}}{m}\left(1 - \frac{p^2}{2m^2c^2}\right) - \frac{e\hbar}{2m^2c^2}\mathbf{S} \times \mathcal{E}. \quad (29)$$

The last term describes topological spin transport of the electron and causes a spin current orthogonal to the electric field, exactly like in spin Hall effect [11–14,21]. As it is shown in [20] this term is precisely two times larger in magnitude then the analogous one that follows from the Pauli Hamiltonian with spin-orbit interaction.[2] In the ultra-relativistic limit the motion equation with an accuracy of $1/p^2$ has the form

$$\dot{\mathbf{r}} = c\frac{\mathbf{p}}{p}\left(1 - \frac{m^2c^2}{2p^2}\right) - \lambda e\hbar \frac{\mathbf{p} \times \mathcal{E}}{p^3}, \quad (30)$$

where $\lambda = (\mathbf{Sp})/2p$ is the electron helicity. Here the topological term has the form of Lorentz force caused by 'magnetic monopole' at the origin of the momentum space. It also describes topological spin transport of spin Hall effect type [11–14,21] and similar to optical Magnus effect [17–20,22]. Analogous equation (with $m = 0$) takes place for photons [17–20], its consequences were verified experimentally [22].

---

[2] Note that the trajectory equations proposed in [9] also lead to the incorrect coefficient 3/8 instead of 1/2 at the last term of Eq. (29).

In a uniform magnetic field (without electric field) a non-relativistic electron moves along the Larmor orbit. We assume $\mathbf{p}.\mathcal{H} = 0$ for simplicity and $\mathbf{S}.\mathcal{H} = \mu.\mathcal{H}$. Then the first Eq. (28) with Eq. (25) with an accuracy of $1/m^3c^3$ yields

$$\dot{\mathbf{p}} = \frac{e}{mc}\left(1 - \frac{p^2}{2m^2c^2} - \mu\frac{e\hbar.\mathcal{H}}{2m^2c^3}\right)\mathbf{p} \times \mathcal{H} + \frac{e}{c}\partial_{\mathbf{p}}\Delta E \times \mathcal{H}. \quad (31)$$

The similar equation that follows from the refined Pauli Hamiltonian (19) (see [26]) has the opposite sign before term with $\mu$. Thus Eq. (31) predicts a splitting of trajectories of two spin states in a way opposite to traditional motion equations. Expression in parentheses in Eq. (31) describes shift of the cyclotron frequency, which can be represented as $\omega_c = \omega_{c0}\left(1 - \frac{p^2}{2m^2c^2} - \frac{e\hbar}{c}\mathbf{F}\mathcal{H}\right)$ with $\omega_{c0} = e.\mathcal{H}/mc$. Due to the distinction in signs at the term with $\mu$ in Eq. (31), our equations predict the cyclotron frequency to be lower (for positive $\mu$) than Pauli-Hamiltonian cyclotron frequency; the difference equals $\Delta\omega_{c0} = \mu\frac{e\hbar.\mathcal{H}}{m^2c^3}$. The ultra-relativistic limit of an analogous problem with an accuracy of $1/p^2$ leads to the motion equations

$$\dot{\mathbf{p}} = \frac{e}{c}\dot{\mathbf{r}} \times \mathcal{H}, \quad \dot{\mathbf{r}} = c\frac{\mathbf{p}}{p}\left(1 - \frac{m^2c^2}{2p^2}\right) + \partial_{\mathbf{p}}\Delta E - \lambda\frac{e\hbar}{p^2}\mathcal{H}. \quad (32)$$

The last term here contains a term that is proportional to electron helicity and is directed along $\mathcal{H}$. It causes topological spin transport of the electron similar to spin Hall effect and optical Magnus effect but in the absence of electric field.

All predictions mentioned above could be experimentally verified in the scattering of relativistic particles in an external electromagnetic field and by measure of the cyclotron frequency shift in a magnetic field.

## VI. CONCLUSION

We have considered semiclassical evolution of an electron in an external electromagnetic field. The Dirac equation has been diagonalized in the first-order approximation in $\hbar$ with an arbitrary particle energy and an arbitrary external field.[3] We have determined the covariant momentum and coordinates for electron [27] and computed their non-trivial commutators, taking into account the Berry gauge potential induced during the diagonalization of the Dirac equation. The commutators contain $U(2)$ gauge field, whose potential is the sum of $U(1)$ electromagnetic potential and $SU(2)$ non-Abelian Berry gauge potential. The main results of this paper are the electron's semiclassical motion equations: polarization evolution equation (which gives rise to BMT equation for the spin precession [7–9]) and Hamiltonian equations for its center's trajectory. These equations effectively contain topological terms defined by the Berry gauge field in the momentum space. Trajectory equations have a covariant form and substantially differ from the Pauli-Hamiltonian equations for canonical variables. We have briefly noted the basic simplest corollaries and new effects that follow from these equations, and shown the correspondence between limit cases and the latest research in topological spin transport of relativistic particles.

---

[3] We considered semiclassical and adiabatic approximations as a formal asymptotics $\hbar \to 0$. In real physical situation these approximations require the smallness of a dimensionless parameter proportional to the wavelength $\hbar/p$ and to the electromagnetic field strength $\frac{e}{c}\max(\mathcal{E},\mathcal{H})$. It can be determined from the requirement that the wavelength must be much smaller than all other characteristic scales of the system evolution and the field strength must be small enough to avoid creation of electron-positron pairs.

The work was partially supported by INTAS (Grant No. 03-55-1921) and by Ukrainian President's Grant for Young Scientists GP/F8/51.

*Note added:* A related paper, D. Xiao, J. Shi, and Q. Niu, cond-mat/0502340, that consider electron in a solid have appeared after the submission of this work. Despite that the authors there use another (possibly more apt) idea about correction to the electron density of states, they derive the commutation relations similar to Eq. (23) and predict analogous shift of the electron's cyclotron frequency.